\documentclass[journal,twoside]{IEEEtran}
\usepackage{graphicx,amsmath,amssymb,color}
\usepackage{verbatim}
\usepackage[noadjust]{cite}
\usepackage{array}
\usepackage{cases}
\usepackage{url}
\usepackage{multirow}
\newcommand{\ket}[1]{\left\vert #1 \right\rangle}

\newtheorem{theorem}{Theorem}[section]
\newtheorem{proposition}[theorem]{Proposition}

\newtheorem{lemma}[theorem]{Lemma}

\begin{document}
\title{Quantum Synchronizable Codes from Finite Geometries}
\author{Yuichiro~Fujiwara,~\IEEEmembership{Member,~IEEE} and Peter Vandendriessche,~\IEEEmembership{Member,~IEEE}%
\thanks{This work was supported by JSPS (Y.F.) and FWO (P.V.).
The second author is support by a PhD fellowship of the Research Foundation - Flanders (FWO).}%
\thanks{Y. Fujiwara is with the Division of Physics, Mathematics and Astronomy, California Institute of Technology, MC 253-37, Pasadena, CA 91125 USA
{(email: yuichiro.fujiwara@caltech.edu)}.}%
\thanks{P. Vandendriessche is with the Department of Mathematics, Ghent University, Krijgslaan 281 - S22, 9000 Ghent, Belgium
{(email: pv@cage.ugent.be)}.}%
\thanks{Copyright \copyright\ 2014 IEEE. Personal use of this material is permitted.
However, permission to use this material for any other purposes must be obtained from the IEEE by sending a request to pubs-permissions@ieee.org.}
}
\markboth{IEEE transactions on Information Theory,~Vol.~x, No.~xx,~month~year}
{Fujiwara and Vandendriessche: Quantum synchronizable codes from finite geometries}

\maketitle

\begin{abstract}
Quantum synchronizable error-correcting codes are special quantum error-correcting codes
that are designed to correct both the effect of quantum noise on qubits and misalignment in block synchronization.
It is known that in principle such a code can be constructed through a combination of a classical linear code and its subcode
if the two are both cyclic and dual-containing.
However, finding such classical codes that lead to promising quantum synchronizable error-correcting codes is not a trivial task.
In fact, although there are two families of classical codes that are proved to produce quantum synchronizable codes
with good minimum distances and highest possible tolerance against misalignment,
their code lengths have been restricted to primes and Mersenne numbers.
In this paper, examining the incidence vectors of projective spaces over the finite fields of characteristic $2$,
we give quantum synchronizable codes from cyclic codes whose lengths are not primes or Mersenne numbers.
These projective geometric codes achieve good performance in quantum error correction
and possess the best possible ability to recover synchronization, thereby enriching the variety of good quantum synchronizable codes.
We also extend the current knowledge of cyclic codes in classical coding theory
by explicitly giving generator polynomials of the finite geometric codes and completely characterizing the minimum weight nonzero codewords.
In addition to the codes based on projective spaces,
we carry out a similar analysis on the well-known cyclic codes
from Euclidean spaces that are known to be majority logic decodable
and determine their exact minimum distances.
\end{abstract}


\IEEEpeerreviewmaketitle

\section{Introduction}\label{intro}
\IEEEPARstart{Q}{uantum} error correction is a crucial element of quantum information science
that plays a key role in realizing quantum information processing in a noisy environment.
Active quantum error correction is an important and extensively studied method for suppressing quantum noise,
where one extracts the information about what types of errors occurred on which qubits
through syndrome extraction without disturbing the quantum information carried by qubits.
Once this information is obtained, the effect of quantum noise may be nullified by applying appropriate quantum operations on affected qubits.

In the context of quantum error correction,
quantum noise is most typically described by operators that act on qubits.
The most general error model of this kind is the linear combinations of the Pauli operators $I$, $X$, $Y$, and $Z$ acting on each qubit \cite{Knill:2000}.
From this point of view, quantum error-correcting codes are schemes that allow for recovering the original quantum state
when unintended operators may act on some qubits.

This type of typical error model may be considered a quantum version of \textit{additive noise},
which is among the most important and well-studied kinds of error models in information theory.
However, it is not the only error model of importance.

An example of errors that do not fall into the category of additive noise but are crucial in information theory is synchronization errors.
The simplest type of synchronization error is \textit{misalignment} with respect to the frame structure of a data stream.
To describe misalignment in the context of quantum information,
assume that we have three qubits $q_0$, $q_1$, $q_2$
and encode each of them by the perfect $5$-qubit code \cite{Laflamme:1996,Bennett:1996}.
Then the quantum information we have can be expressed by a sequence of fifteen qubits,
where each $5$-qubit state $\ket{\psi_i}$, $i=0, 1, 2$, represents one logical qubit of quantum information that corresponds to the original qubit $q_i$.
In order to correctly process quantum information,
we need to know the exact location of the boundary of each $5$-qubit block in the $15$-qubit state $\ket{\psi_0}\ket{\psi_1}\ket{\psi_2}$.
For instance, if misalignment occurs by two qubits to the left when handling the stream of fifteen qubits,
our quantum error correction device trying to correct errors on $\ket{\psi_1}$
will apply the quantum operation on the wrong set of five qubits, two of which come from $\ket{\psi_0}$ and three of which belong to $\ket{\psi_1}$.

In classical coding theory, error-correcting codes that can correct both additive noise and misalignment in block synchronization
are called \textit{synchronizable error-correcting codes} \cite{Bose:1967}.
This paper studies the quantum analogue of such active error correction schemes that allow for extracting the information about
the magnitude and direction of misalignment through measurement
while simultaneously identifying the types and positions of standard quantum errors that may have occurred on qubits.

Formally speaking, we call a coding scheme a \textit{quantum synchronizable} $(a_l, a_r)$-$[[n,k]]$ \textit{code}
if it encodes $k$ logical qubits into $n$ physical qubits and
corrects misalignment by up to $a_l$ qubits to the left and up to $a_r$ qubits to the right.
In order to suppress as diverse quantum noise as possible,
we would also like our quantum synchronizable codes to be able to correct linear combinations of $I$, $X$, $Z$, and $Y$ that act on physical qubits.

While it may appear quite difficult to devise an efficient and error-tolerant scheme that also corrects misalignment in the context of quantum information,
it has been proved that in principle a known quantum error correction technique can be extended to the case of block synchronization recovery.
The first examples of quantum synchronizable codes with standard quantum error correction capabilities were presented in \cite{Fujiwara:2013},
where a general framework for code construction as well as their encoding and decoding methods were developed.
This framework was subsequently improved by a more extensive use of finite algebra in \cite{Fujiwara:2013d}.
In both cases, classical error-correcting codes with special algebraic properties are exploited
in a way similar to the well-known Calderbank-Shor-Steane (CSS) construction \cite{Calderbank:1996,Steane:1996,Steane:1996a}.

However, these first theoretical steps towards solving the problem of block synchronization for quantum information left many challenges as well.
One of the main hurdles in the theoretical study of quantum synchronizable codes
is that it is quite difficult to find suitable classical error-correcting codes
because the required algebraic constraints are very severe and difficult to analyze.
In fact, there are only two known classes of classical error-correcting codes that are proved to possess the required properties
while achieving good quantum error correction capabilities and high tolerance against misalignment at the same time.

Particularly constrained is the variety of available code parameters.
For instance, the lengths of the encoded information blocks of
the known quantum synchronizable codes that have highest possible tolerance against misalignment
are all primes or of the form $2^m-1$, that is, Mersenne numbers.

The primary purpose of this paper is to enrich the spectrum of quantum synchronizable codes
by giving an infinite family in which the encoded block lengths can be neither primes nor Mersenne numbers.
We prove that these codes have the highest possible tolerance against misalignment
and are capable of correcting the effect of a substantial level of standard quantum noise.
To this end, we use the theory of finite geometries
and introduce a class of classical error-correcting codes suitable for use as the ingredients of quantum synchronizable codes.

As a by-product of our analysis, we also extend the current knowledge of classical coding theory.
We exploit one of the most important classes of classical error-correcting codes, namely \textit{cyclic codes} \cite{Pless:1998}.
While cyclic codes are of practical and theoretical importance,
their true minimum distances are notoriously difficult to compute.
In the study that follows, we show various properties of cyclic codes based on two types of finite geometries, that is,
projective geometry and Euclidean geometry.
For the case of projective geometry, not only do we prove the exact minimum distances,
but we also completely characterize the nonzero codewords of minimum weight.
A similar analysis is carried out for a well-known class of cyclic codes from Euclidean geometry which have extensively been studied
both in classical coding theory and in quantum coding theory.
We determine the exact minimum distances of the Euclidean cyclic codes
and also prove that they are capable of recovering synchronization from severe misalignment if used as quantum synchronizable codes.

In the next section we briefly review the improved framework of quantum synchronizable codes given in \cite{Fujiwara:2013d}
from the viewpoint of what kind of classical error-correcting code is required.
Section \ref{main} examines special structures of finite geometries
and shows that they form suitable classical error-correcting codes for our purpose.
Concluding remarks are given in Section \ref{conclusion}.

\section{General construction}\label{framework}
This section describes the properties of classical error-correcting codes required for constructing quantum synchronizable error-correcting codes.
The proofs of the basic facts in coding theory we use in this paper can be found in a standard textbook such as \cite{MacWilliams:1977}.
For mathematical details of the problem of block synchronization for quantum information,
we refer the reader to \cite{Fujiwara:2013d,Fujiwara:2013i}.

A \textit{binary linear} $[n,k,d]$ {code} of \textit{length} $n$, \textit{dimension} $k$, and \textit{minimum distance} $d$
is a $k$-dimensional subspace $\mathcal{L}$ of the $n$-dimensional vector space $\mathbb{F}_2^n$ over the finite field $\mathbb{F}_2$ of order $2$
such that $\min\left\{\text{wt}(\boldsymbol{v}) \mid \boldsymbol{v} \in \mathcal{L}\setminus\{\boldsymbol{0}\}\right\} = d$,
where $\text{wt}(\boldsymbol{v})$ is the number of nonzero coordinates of $\boldsymbol{v}$.
Let $\mathcal{C}$ and $\mathcal{D}$ be two binary linear codes of the same length.
$\mathcal{D}$ is $\mathcal{C}$-\textit{containing} if $\mathcal{C} \subseteq \mathcal{D}$.
It is \textit{dual-containing} if it contains its dual $\mathcal{D}^{\perp} = \bigl\{\boldsymbol{d}^{\perp} \in \mathbb{F}_2^n \mid
\boldsymbol{d} \cdot \boldsymbol{d}^{\perp} = \boldsymbol{0}, \boldsymbol{d} \in \mathcal{D}\bigr\}$.
In what follows, we always assume that classical codes are over $\mathbb{F}_2$ and omit the term binary.

A \textit{cyclic} $[n,k,d]$ code $\mathcal{C}$ is a linear $[n,k,d]$ code in which
every cyclic shift of every codeword $\boldsymbol{c} \in \mathcal{C}$ is also a codeword,
that is, for any $\boldsymbol{c}  = (c_0,\dots, c_{n-1}) \in \mathcal{C}$, we have $(c_1,\dots, c_{n-1}, c_0) \in \mathcal{C}$.
It is known that, by regarding each codeword as the coefficient vector of a polynomial in $\mathbb{F}_2[x]$,
a cyclic code of length $n$ can be seen as a principal ideal in the ring $\mathbb{F}_2[x]/(x^n-1)$
generated by the unique monic nonzero polynomial $g(x)$ of minimum degree in the code which divides $x^n-1$.
When a cyclic code is of length $n$ and dimension $k$,
the set of codewords can be written as $\mathcal{C} = \left\{i(x)g(x) \mid \deg(i(x)) < k\right\}$, where the degree $\deg(g(x))$ of the generator polynomial is $n-k$.

The improved framework for constructing quantum synchronizable codes involves an algebraic notion in polynomial rings.
Let $f(x) \in \mathbb{F}_2[x]$ be a polynomial over $\mathbb{F}_2$ such that $f(0)=1$.
The cardinality $\operatorname{ord}(f(x)) = \left\vert\left\{x^a \pmod{f(x)} \mid a \in \mathbb{N}\right\}\right\vert$ is called the \textit{order} of the polynomial $f(x)$,
where $\mathbb{N}$ is the set of positive integers.
This cardinality is also known as the \textit{period} or \textit{exponent} of $f(x)$ in the literature.
The following is the improved general construction stated in the form of a mathematical theorem.

\begin{theorem}[\cite{Fujiwara:2013d}]\label{knowntheorem}
Let $\mathcal{C}$ be a dual-containing cyclic $[n, k_1,d_1]$ code
with generator polynomial $h(x)$ and $\mathcal{D}$ a $\mathcal{C}$-containing cyclic $[n, k_2, d_2]$ code with generator polynomial $g(x)$.
Define polynomial $f(x)$ of degree $k_2 - k_1$ to be the quotient of $h(x) = f(x)g(x)$ divided by $g(x)$ over $\mathbb{F}_2[x]/(x^n-1)$.
For every pair $a_l$, $a_r$ of nonnegative integers such that $a_l + a_r < \operatorname{ord}(f(x))$
there exists a quantum synchronizable $(a_l, a_r)$-$[[n+a_l+a_r, 2k_1-n]]$ code that corrects
at least up to $\left\lfloor\frac{d_1-1}{2}\right\rfloor$ phase errors and
at least up to $\left\lfloor\frac{d_2-1}{2}\right\rfloor$ bit errors.
\end{theorem}

For the proof of the theorem above and the procedures for encoding and decoding,
the reader is referred to \cite{Fujiwara:2013d,Fujiwara:2013}.

From the viewpoint of searching for suitable classical error-correcting codes,
an important fact to note is that if a linear code $\mathcal{C}$ is dual-containing, a $\mathcal{C}$-containing linear code is also dual-containing.
Hence, what Theorem \ref{knowntheorem} actually requires is a pair of dual-containing cyclic codes, one of which is contained in the other
and both of which guarantee large minimum distances.
In addition to being nested, dual-containing, cyclic, and of large minimum distance,
it is desirable for the pair of linear codes to lead to as large $\operatorname{ord}(f(x))$
as possible in order to tolerate the widest range of misalignment.
Note that for any pair of cyclic codes of length $n$, the corresponding value of $\operatorname{ord}(f(x))$ is at most $n$.
This is because $f(x)$ is a divisor of the generator polynomial of the smaller cyclic code.
In other words, $f(x)$ also divides $x^n-1$, so that $x^{a} = x^{a+n} \pmod{f(x)}$ for any integer $a$.

The known quantum synchronizable codes employ well-known classes of cyclic codes
called \textit{narrow-sense Bose-Chaudhuri-Hocquenghem} (BCH) \textit{codes} and \textit{punctured Reed-Muller codes}.
Their precise definitions, parameters, and other important facts in the context of quantum block synchronization
can be found in \cite{Fujiwara:2013d} and references therein.
The notable point is that these cyclic codes have substantial minimum distances
while being both dual-containing and nested if their dimensions are large enough.
It can be shown that the corresponding orders $\operatorname{ord}(f(x))$ often takes the maximum value.
However, the lengths of the codes of the former class which has been studied
for synchronization recovery to a substantial depth are all of the form $2^m-1$ or primes.
The latter class only contains codes of length that is simultaneously prime and of the form $2^m-1$.
The goal of the next section is to prove that these are not the only suitable error-correcting codes
by giving explicit examples whose lengths have not be realized by the previously known families.

\section{Subspaces of finite geometries and quantum synchronizable codes}\label{main}
This section examines properties of finite geometries and provide a family of quantum synchronizable codes.
The proofs of the basic facts and notions regarding finite geometries we need can be found in \cite{Hirschfeld:1998}.

We divide this section into two subsections.
Section \ref{projective} studies codes based on projective geometry.
Codes based on Euclidean geometry are examined in Section \ref{euclidean}.
In both cases, the lengths, dimensions, and minimum distances of our classical error-correcting codes are precisely and theoretically determined.
We also prove that the maximum tolerable magnitudes $\operatorname{ord}(f(x))$ of misalignment of the corresponding quantum synchronizable codes
are always the same as their code lengths, which are the highest possible values.

\subsection{Projective geometry codes}\label{projective}
Let $m$, $h$, and $t$ be positive integers such that $t \leq m-1$.
The \textit{projective geometry} $\textup{PG}(m, 2^h)$ of \textit{dimension} $m$ over $\mathbb{F}_{2^h}$
is a finite geometry whose \textit{points} and $t$-\textit{dimensional subspaces} are the $1$-dimensional vector subspaces
and the $(t+1)$-dimensional vector subspaces of the $(m+1)$-dimensional vector space $\mathbb{F}_{2^h}^{m+1}$ respectively.
Because the points are the $1$-dimensional vector subspaces,
the set $P$ of points in $\textup{PG}(m, 2^h)$ is of cardinality $\frac{2^{h(m+1)}-1}{2^h-1}$.

Take a $t$-dimensional subspace $\pi$ of $\textup{PG}(m, 2^h)$.
The \textit{incidence vector} $\chi_{\pi}$ of $\pi$ is the binary $\frac{2^{h(m+1)}-1}{2^h-1}$-dimensional vector such that
the coordinates are indexed by the points and such that
each entry is $1$ if $\pi$ contains the corresponding point and $0$ otherwise.
Let $\mathcal{B}$ be the set of $t$-dimensional subspaces of $\textup{PG}(m, 2^h)$.
It is known that there exists a collineation $\sigma$ such that
the group $\langle \sigma \rangle$ of order $\frac{2^{h(m+1)}-1}{2^h-1}$ acts regularly on the points in $\textup{PG}(m, 2^h)$ \cite{Singer:1938},
which means that the incidence relation in the set system $(P, \mathcal{B})$ is invariant under the action of the cyclic group $\langle \sigma \rangle$.
Therefore, by regarding points as the elements of the cyclic group and indexing the coordinates of each incidence vector accordingly
by $\sigma^0, \sigma^1, \sigma,^2, \dots, \sigma^\frac{2^{h(m+1)}-1}{2^h-1}$ in the natural order,
for any incidence vector
\[\chi_{\pi} = (x_0, \dots, x_{\frac{2^{h(m+1)}-1}{2^h-1}-1})\]
of a $t$-dimensional subspace $\pi \in \mathcal{B}$,
its cyclic shift
\[(x_1, \dots, x_{\frac{2^{h(m+1)}-1}{2^h-1}-1}, x_0)\]
is also the incidence vector of some $t$-dimensional subspace.
Hence, the vector space $\mathcal{P}_{m,t,2^h} = \langle \chi_\pi \mid \pi \in \mathcal{B} \rangle$
spanned by the incidence vectors of $t$-dimensional subspaces in $\textup{PG}(m, 2^h)$
can be seen as a cyclic code of length $\frac{2^{h(m+1)}-1}{2^h-1}$.
We assume that the coordinates are indexed by the points of $\textup{PG}(m, 2^h)$ in this cyclic way throughout this subsection.

The \textit{complement} $\chi_{\overline{\pi}}$ of an incidence vector $\chi_{\pi}$ is the vector
$\chi_{\overline{\pi}} = \chi_{\pi}+\boldsymbol{1}$, where $\boldsymbol{1}$ is the all-one vector.
In other words, $\chi_{\overline{\pi}}$ is obtained by flipping all $0$'s and $1$'s in $\chi_{\pi}$.
Let $\mathcal{C}_{m,t,2^h} = \langle \chi_{\overline{\pi}} \mid \pi \in \mathcal{B} \rangle^\perp$ be
the dual of the vector space spanned by the set of complements of the incidence vectors of $t$-dimensional subspaces of $\textup{PG}(m, 2^h)$.
We prove that $\mathcal{C}_{m,t,2^h}$ satisfies the required conditions for use as ingredients of quantum synchronizable codes
while achieving good quantum error correction capabilities and high tolerance against misalignment
if $t$ is in a suitable range with respect to the size of $m$.
More specifically, we show the following theorem.
\begin{theorem}\label{PGmain}
Let $m$, $h$, and $t$ be positive integers such that $\frac{m+1}{2} \leq t \leq m-2$.
For every pair $a_l$, $a_r$ of nonnegative integers such that $a_l + a_r < \frac{2^{h(m+1)}-1}{2^h-1}$
there exists a quantum synchronizable $(a_l, a_r)$-$\left[\left[\frac{2^{h(m+1)}-1}{2^h-1}+a_l+a_r,
\frac{2^{h(m+1)}-1}{2^h-1}-2\dim\mathcal{P}_{m,t,2^h}+2\right]\right]$ code that corrects
at least up to $\frac{2^{h(m-t+1)-1}-2^{h-1}}{2^h-1}$ phase errors and
at least up to $\frac{2^{h(m-t)-1}-2^{h-1}}{2^h-1}$ bit errors.
\end{theorem}

As we have seen in the previous section, Theorem \ref{knowntheorem} requires classical error-correcting codes to simultaneously satisfy various properties.
The fact that $\mathcal{C}_{m,t,2^h} = \langle \chi_{\overline{\pi}} \mid \pi \in \mathcal{B} \rangle^\perp$ is a cyclic code
follows directly from the fact that $\mathcal{P}_{m,t,2^h} = \langle \chi_\pi \mid \pi \in \mathcal{B} \rangle$ is cyclic as a linear code.
\begin{proposition}\label{PGcyclic}
$\mathcal{C}_{m,t,2^h}$ is cyclic as a linear code.
\end{proposition}

To form a quantum synchronizable code, a cyclic code must be dual-containing.
\begin{lemma}\label{PGdualcontaining}
If $\frac{m+1}{2} \leq t \leq m-1$, then $\mathcal{C}_{m,t,2^h}^\perp \subseteq \mathcal{C}_{m,t,2^h}$.
\end{lemma}
\begin{IEEEproof}
It suffices to show that for any pair $\overline{\pi_0}, \overline{\pi_1} \in \mathcal{B}$ of $t$-dimensional subspaces of $\textup{PG}(m, 2^h)$,
the corresponding incidence vectors $\chi_{\overline{\pi_0}}, \chi_{\overline{\pi_1}} \in \mathcal{C}_{m,t,2^h}^\perp$ are orthogonal to each other.
Note that $\chi_{\overline{\pi_0}}$ and $\chi_{\overline{\pi_1}}$ are orthogonal to each other if and only if
the cardinality $\left\vert \overline{\pi_0} \cap \overline{\pi_1} \right\vert$ is even.
Note also that
\begin{align*}
\left\vert \overline{\pi_0} \cap \overline{\pi_1} \right\vert &= \frac{2^{h(m+1)}-1}{2^h-1} - \left\vert \pi_0 \cup \pi_1 \right\vert\\
&=  \frac{2^{h(m+1)}-1}{2^h-1} - \vert \pi_0 \vert - \vert \pi_1 \vert + \vert \pi_0 \cap \pi_1 \vert.
\end{align*}
If $t \geq \frac{m+1}{2}$, the intersection $\pi_0 \cap \pi_1$ is a nonempty subspace of $\textup{PG}(m, 2^h)$.
Because any nonempty subspace of $\textup{PG}(m, 2^h)$ contains an odd number of points,
the four terms $\frac{2^{h(m+1)}-1}{2^h-1}$, $\vert \pi_0 \vert$, $\vert \pi_1 \vert$, and $\vert \pi_0 \cap \pi_1 \vert$ are all odd.
Thus, their sum, whose parity is the same as that of $\left\vert \overline{\pi_0} \cap \overline{\pi_1} \right\vert$, is indeed even.
\end{IEEEproof}

The following lemma shows that $\mathcal{C}_{m,t,2^h}$ has a suitable nested property.
\begin{lemma}\label{PGnest}
If $2 \leq t \leq m-1$, then $\mathcal{C}_{m,t-1,2^h} \subset \mathcal{C}_{m,t,2^h}$.
\end{lemma}

Note that recursively applying the above lemma shows that $\mathcal{C}_{m,t,2^h}$ contains $\mathcal{C}_{m,t',2^h}$ for all $t' \leq t$. 
To prove Lemma \ref{PGnest}, we use two simple facts about our cyclic codes.
\begin{proposition}\label{prop1}
$\mathcal{P}_{m,t,2^h} = \langle \mathcal{C}_{m,t,2^h}^\perp, \boldsymbol{1} \rangle$, where
the $\frac{2^{h(m+1)}-1}{2^h-1}$-dimensional all-one vector $\boldsymbol{1} \not\in \mathcal{C}_{m,t,2^h}^\perp$
\end{proposition}
\begin{IEEEproof}
Because all generators $\chi_{\overline{\pi}}$ of $\mathcal{C}_{m,t,2^h}^\perp$ are of even weight,
all codewords of $\mathcal{C}_{m,t,2^h}^\perp$ are also of even weight.
Since the length $\frac{2^{h(m+1)}-1}{2^h-1}$ of this cyclic code is odd,
the all-one vector is not a codeword.
Recall that $\mathcal{P}_{m,t,2^h}$ is the vector space
spanned by the incidence vectors of $t$-dimensional subspaces in $\textup{PG}(m, 2^h)$.
Because the number of $t$-dimensional subspaces that contain a given point in $\textup{PG}(m, 2^h)$ is always odd,
we have
\[\sum_{\pi \in \mathcal{B}}\chi_{\pi} = \boldsymbol{1},\]
which implies that $\boldsymbol{1} \in \mathcal{P}_{m,t,2^h}$.
Because $\chi_{\overline{\pi}} = \chi_{\pi}+\boldsymbol{1}$,
it follows that $\mathcal{C}_{m,t,2^h}^\perp \subset \mathcal{P}_{m,t,2^h}$.
Thus, we have $\langle \mathcal{C}_{m,t,2^h}^\perp, \boldsymbol{1} \rangle \subseteq \mathcal{P}_{m,t,2^h}$.
Now the fact that $\chi_{\overline{\pi}} = \chi_{\pi}+\boldsymbol{1}$ is equivalent to the relation that $\chi_{\pi} = \chi_{\overline{\pi}}+\boldsymbol{1}$,
which implies that $\mathcal{P}_{m,t,2^h} \subseteq \langle \mathcal{C}_{m,t,2^h}^\perp, \boldsymbol{1} \rangle$.
Thus, $\mathcal{P}_{m,t,2^h} = \langle \mathcal{C}_{m,t,2^h}^\perp, \boldsymbol{1} \rangle$ as desired.
\end{IEEEproof}

\begin{proposition}\label{prop2}
$\mathcal{C}_{m,t,2^h} = \langle\mathcal{P}_{m,t,2^h}^\perp, \boldsymbol{1}\rangle$,
where $\boldsymbol{1} \not\in \mathcal{P}_{m,t,2^h}^\perp$.
\end{proposition}
\begin{IEEEproof}
Because the generators of $\mathcal{P}_{m,t,2^h}$ are all of odd weight,
the inner product between $\boldsymbol{1}$ and any of the generators is nonzero,
which implies that $\boldsymbol{1} \not\in \mathcal{P}_{m,t,2^h}^\perp$.
By the same token, because the generators of $\mathcal{C}_{m,t,2^h}^\perp$ are of even weight,
we have $\boldsymbol{1} \in \mathcal{C}_{m,t,2^h}$.
By Proposition \ref{prop1}, $\mathcal{C}_{m,t,2^h}^\perp \subset \mathcal{P}_{m,t,2^h}$,
which implies that $\mathcal{P}_{m,t,2^h}^\perp \subset \mathcal{C}_{m,t,2^h}$.
Again by Proposition \ref{prop1}, the dimensions of $\mathcal{P}_{m,t,2^h}$ and $\mathcal{C}_{m,t,2^h}^\perp$
satisfy the equation that $\dim \mathcal{P}_{m,t,2^h} = \dim \mathcal{C}_{m,t,2^h}^\perp +1$.
Hence, $\mathcal{C}_{m,t,2^h} = \langle\mathcal{P}_{m,t,2^h}^\perp, \boldsymbol{1}\rangle$ as desired.
\end{IEEEproof}

\begin{IEEEproof}[Proof of Lemma \ref{PGnest}]
Assume that $2 \leq t \leq m-1$.
Take a $t$-dimensional subspace $\pi$ of $\textup{PG}(m,2^h)$ and
define $\Pi$ to be the set of $(t-1)$-dimensional subspaces that are contained in $\pi$.
The set $\Pi$ contains exactly $\frac{2^{ht}-1}{2^h-1}$ $(t-1)$-subspaces containing a given point $p \in \pi$.
Since this number is odd, we have
\[\chi_{\pi} = \sum_{\pi' \in \Pi}\chi_{\pi'}.\]
Thus, $\mathcal{P}_{m,t,2^h} \subset \mathcal{P}_{m,t-1,2^h}$, which is equivalent to $\mathcal{P}_{m,t-1,2^h}^\perp \subset \mathcal{P}_{m,t,2^h}^\perp$.
Hence, by Proposition \ref{prop2}, we have $\mathcal{C}_{m,t-1,2^h} \subset \mathcal{C}_{m,t,2^h}$.
The proof is complete.
\end{IEEEproof}

Proposition \ref{PGcyclic} and Lemmas \ref{PGdualcontaining} and \ref{PGnest} together
prove that for any pair $t$, $t'$ of positive integers such that $2 \leq t' < t \leq m-1$,
the pair $\mathcal{C}_{m,t,2^h}$, $\mathcal{C}_{m,t',2^h}$ of linear codes satisfy the three conditions of cyclicity, duality, and nestedness
required to construct a quantum synchronizable code through Theorem \ref{knowntheorem}.
Because we already know that their length is $\frac{2^{h(m+1)}-1}{2^h-1}$,
the remaining task is to determine their dimensions, minimum distances, and the maximum tolerable magnitude of misalignment.

The dimension of $\mathcal{C}_{m,t,2^h}$ is determined by that of $\mathcal{P}_{m,t,2^h}$.
\begin{lemma}\label{PGdim}
For positive integers $m$, $h$, $t$ such that $t \leq m-1$,
\[\dim\mathcal{C}_{m,t,2^h} = \frac{2^{h(m+1)}-1}{2^h-1} - \dim\mathcal{P}_{m,t,2^h}+1.\]
\end{lemma}
\begin{IEEEproof}
By Proposition \ref{prop2}, we have
\begin{align*}
\dim\mathcal{C}_{m,t,2^h} &= \dim\mathcal{P}_{m,t,2^h}^\perp+1\\
&=  \min\left\{\vert P\vert, \vert\mathcal{B}\vert\right\} - \dim\mathcal{P}_{m,t,2^h}+1,
\end{align*}
where $P$ is the set of points and $\mathcal{B}$ is the set of $t$-dimensional subspaces of $\textup{PG}(m, 2^h)$.
The number $\frac{2^{h(m+1)}-1}{2^h-1}$ of points is always smaller than or equal to that of $t$-dimensional subspaces.
\end{IEEEproof}

The following well-known formula gives the exact value of $\dim\mathcal{P}_{m,t,2^h}$,
allowing for calculating the dimensions of our cyclic codes.
\begin{theorem}[\cite{Hamada:1973}]\label{HamadaPGdimension}
For positive integers $m$, $h$, $t$ such that $t \leq m-1$,
\begin{align*}
\dim\mathcal{P}_{m,t,2^h} =
\sum_{(s_0,s_1,\dots,s_h)}&\prod_{j=0}^{h-1} \sum_{i=0}^{\left\lfloor\frac{2s_{j+1}-s_j}{2}\right\rfloor} (-1)^i \binom{m+1}{i}\\
&\times\binom{m+2s_{j+1}-s_j-2i}{m},
\end{align*}
where the first summation is taken over all $(s_0,s_1,\dots,s_h)$ with
$s_h=s_0$; $t+1\leq s_j\leq m+1$ for all $j$ with $0\leq j<h$, and $0\leq 2s_{j+1}-s_j\leq m+1$ for all $j$ with $0\leq j<h$. 
\end{theorem}

To examine the minimum distance of $\mathcal{C}_{m,t,2^h}$, we employ the following two theorems.
\begin{theorem}[\cite{Calkin:1999}]\label{min1}
The minimum distance of $\mathcal{P}_{m,t,2^h}^\perp$ is $(2^h+2)2^{h(m-t-1)}$.
\end{theorem}
\begin{theorem}[\cite{Limbupasiriporn:2012}]\label{min2}
The codewords of $\mathcal{P}_{m,t,2^h}^\perp$ that have the largest number of nonzero entries
are of weight $\frac{2^{h(m+1)}(1-2^{-ht})}{2^h-1}$.
\end{theorem}

We are now able to give the complete profile of the parameters of $\mathcal{C}_{m,t,2^h}$ as a linear code.
\begin{theorem}\label{PGpara}
$\mathcal{C}_{m,t,2^h}$ is a linear $[\frac{2^{h(m+1)}-1}{2^h-1}, \frac{2^{h(m+1)}-1}{2^h-1}-\dim\mathcal{P}_{m,t,2^h}+1,\frac{2^{h(m-t+1)}-1}{2^h-1}]$ code.
\end{theorem}
\begin{IEEEproof}
It suffices to prove that the minimum distance is exactly $\frac{2^{h(m-t+1)}-1}{2^h-1}$.
By Proposition \ref{prop2}, a codeword of $\mathcal{C}_{m,t,2^h}$ is either contained in $\mathcal{P}_{m,t,2^h}^\perp$
or of the form $\boldsymbol{c} + \boldsymbol{1}$ for some $\boldsymbol{c} \in \mathcal{P}_{m,t,2^h}^\perp$.
By Theorem \ref{min1}, among the codewords of the former kind,
the ones with the smallest number of nonzero entries are of weight exactly $(2^h+2)2^{h(m-t-1)}$.
Among the codewords of the latter kind,
the ones with the smallest number of nonzero entries are those obtained
as the complements of the codewords of $\mathcal{P}_{m,t,2^h}^\perp$ that have the largest number of nonzero entries.
Hence, by Theorem \ref{min2}, the codewords of the latter kind that have the smallest number of nonzero entries are of weight exactly
$\frac{2^{h(m+1)}-1}{2^h-1} - \frac{2^{h(m+1)}(1-2^{-ht})}{2^h-1} = \frac{2^{h(m-t+1)}-1}{2^h-1}$.
For all positive $h$ and positive $t < m$, we have $\frac{2^{h(m-t+1)}-1}{2^h-1} < (2^h+2)2^{h(m-t-1)}$.
\end{IEEEproof}

An interesting observation is that the maximum weight codewords in the dual $\mathcal{P}_{m,t,2^h}^\perp$
are the incidence vector of the complement of $(m-t)$-spaces in $\textup{PG}(m,2^h)$ (see \cite[Theorem 3.1]{Limbupasiriporn:2012}).
Thus, the proof above actually shows that the minimum weight nonzero codewords in $\mathcal{C}_{m,t,2^h}$
are exactly the incidence vectors of $(m-t)$-spaces in $\textup{PG}(m,2^h)$,
giving a complete picture of how the nonzero codewords of minimum weight are formed and how many there are.

As shown in Theorem \ref{PGpara}, $\mathcal{C}_{m,t,2^h}$ has a nontrivially large minimum distance and dimension as a cyclic code.
Table \ref{table1} lists the parameters for some $m$, $h$, and $t$.
When $h\not= 1$, the length of $\mathcal{C}_{m,t,2^h}$ is not a Mersenne number and can generally be a composite number,
which has not been realized by previously known cyclic codes that lead to quantum synchronizable codes
with excellent synchronization recovery capabilities.
It should be noted that if $h = 1$, primitive BCH codes of the same length are in general as good or better in terms of dimension and minimum distance.
\begin{table}
\renewcommand{\arraystretch}{1.3}
\caption{Sample parameters of $\mathcal{C}_{m,t,2^h}$ from $\textup{PG}(m,2^h)$.}
\label{table1}
\centering
\begin{tabular}{rrrrrr}
\hline\hline
$m$ & $h$ & $t$ & Length & Dimension & Minimum distance \\\hline
4 & 1 & 2 & 31 & 16 & 7\phantom{0000000}\\
4 & 1 & 3 & 31 & 26 & 3\phantom{0000000}\\
4 & 2 & 2 & 341 & 196 & 21\phantom{0000000}\\
4 & 2 & 3 & 341 & 316 & 5\phantom{0000000}\\
4 & 3 & 2 & 4681 & 3106 & 73\phantom{0000000}\\
4 & 3 & 3 & 4681 & 4556 & 9\phantom{0000000}\\
5 & 1 & 3 & 63 & 42 & 7\phantom{0000000}\\
5 & 1 & 4 & 63 & 57 & 3\phantom{0000000}\\
5 & 2 & 3 & 1365 & 1064 & 21\phantom{0000000}\\
5 & 2 & 4 & 1365 & 1329 & 5\phantom{0000000}\\
5 & 3 & 3 & 37449 & 32598 & 73\phantom{0000000}\\
5 & 3 & 4 & 37449 & 37233 & 9\phantom{0000000}\\
6 & 1 & 3 & 127 & 64 & 15\phantom{0000000}\\
6 & 1 & 4 & 127 & 99 & 7\phantom{0000000}\\
6 & 1 & 5 & 127 & 120 & 3\phantom{0000000}\\
6 & 2 & 3 & 5461 & 3186 & 85\phantom{0000000}\\
6 & 2 & 4 & 5461 & 4901 & 21\phantom{0000000}\\
6 & 2 & 5 & 5461 & 5412 & 5\phantom{0000000}\\
7 & 1 & 4 & 255 & 163 & 15\phantom{0000000}\\
7 & 1 & 5 & 255 & 219 & 7\phantom{0000000}\\
7 & 1 & 6 & 255 & 247 & 3\phantom{0000000}\\
7 & 2 & 4 & 21845 & 16629 & 85\phantom{0000000}\\
7 & 2 & 5 & 21845 & 20885 & 21\phantom{0000000}\\
7 & 2 & 6 & 21845 & 21781 & 5\phantom{0000000}\\
8 & 1 & 4 & 511 & 256 & 31\phantom{0000000}\\
8 & 1 & 5 & 511 & 382 & 15\phantom{0000000}\\
8 & 1 & 6 & 511 & 466 & 7\phantom{0000000}\\
8 & 1 & 7 & 511 & 502 & 3\phantom{0000000}\\
8 & 2 & 4 & 87381 & 51396 & 341\phantom{0000000}\\
8 & 2 & 5 & 87381 & 76512 & 85\phantom{0000000}\\
8 & 2 & 6 & 87381 & 85836 & 21\phantom{0000000}\\
8 & 2 & 7 & 87381 & 87300 & 5\phantom{0000000}\\
\hline\hline
\end{tabular}
\end{table}

The final criterion for being ideal ingredients of quantum synchronizable codes is to be able to provide high tolerance against synchronization errors.
We prove that any pair of cyclic codes in the nested chain
$\mathcal{C}_{m,\lceil\frac{m+1}{2}\rceil,2^h} \subset  \mathcal{C}_{m,\lceil\frac{m+1}{2}\rceil+1,2^h} \subset \dots \subset
\mathcal{C}_{m,m-1,2^h}$ attain the upper bound on the synchronization recovery capabilities.

To investigate the tolerable magnitude of misalignment,
we first determine the generator polynomial of $\mathcal{C}_{m,t,2^h}$
and then apply theorems in finite algebra to explicitly spell out the exact values of $\operatorname{ord}(f(x))$ in Theorem \ref{knowntheorem}
for the case when the cyclic codes are chosen from our nested chain.

To this end, we use the \textit{weight} $\operatorname{w}_{2^h}(a)$ of the $2^h$-ary expansion of a positive integer $a$,
that is,
\[\operatorname{w}_{2^h}(a) = \sum_i a_i,\]
where addition is performed over $\mathbb{Z}$ and
\[a = \sum_{i\in \mathbb{N}\cup\{0\}} a_i2^{hj}\]
with $0 \leq a_i \leq 2^h-1$.
The following theorem gives the explicit form of the generator polynomial of $\mathcal{C}_{m,t,2^h}$.
\begin{theorem}\label{PGgeneratorpoly}
Let $\alpha$ be a primitive element in $\mathbb{F}_{2^{h(m+1)}}$ and $\beta = \alpha^{2^h-1}$.
The generator polynomial $g(x)$ of $\mathcal{C}_{m,t,2^h}$ is
\[g(x) = \prod_{j \in I_{m,t,h}}(x-\beta^j),\]
where
\begin{align*}
I_{m,t,h} = &\left\{ a \in \mathbb{N} \ \middle\vert\ a \leq \frac{2^{h(m+1)}-1}{2^h-1},\right.\\
&\ \left.\max_{0 \leq i \leq h}\operatorname{w}_{2^h}(a(2^h-1)2^i) \leq (m-t)(2^h-1)\right\}.
\end{align*}
\end{theorem}
\begin{IEEEproof}
It is known that $\mathcal{P}_{m,t,2^h}$ is the subfield subcode in $\mathbb{F}_2$ of a punctured generalized Reed-Muller codes \cite{Assmus-Jr.:1998}.
The generator polynomial $h(x)$ of its dual $\mathcal{P}_{m,t,2^h}^\perp$ is
\[h(x) = (x-1)\prod_{j \in I_{m,t,h}}(x-\beta^j)\]
(see \cite[Theorem 13.9.2]{Blahut:2003}\footnote{To avoid confusion in notation,
``$m$'' in Theorem 13.9.2 in \cite{Blahut:2003} corresponds to ``$m+1$'' in this paper while
``$r$'' and ``$s$'' there are ``$t$'' and ``$h$'' here respectively.
Note also that the current edition of the textbook contains typographical errors in the statement of Theorem 13.9.2,
so that ``$0 < j\dots$'' and ``$0 < \max\dots$'' should read ``$0\leq j\dots$'' and ``$0\leq\max\dots$'' respectively.}).
It suffices to show that $h(x)=(x-1)g(x)$.
By Lemma \ref{prop2}, $\mathcal{P}_{m,t,2^h}^\perp \subset \mathcal{C}_{m,t,2^h}$
and $\dim\mathcal{C}_{m,t,2^h} = \dim\mathcal{P}_{m,t,2^h}^\perp+1$.
Thus, the generator polynomial $g(x)$ of $\mathcal{C}_{m,t,2^h}$ is a divisor of $h(x)$,
where the quotient is a polynomial of degree $1$ over $\mathbb{F}_2$.
Since $h(0)=1$, the polynomial $x$ is not a factor of $h(x)$. Hence, we have $h(x)=(x-1)g(x)$ as desired.
\end{IEEEproof}

To show that $\mathcal{C}_{m,t,2^h}$ gives the highest possible tolerance against misalignment,
we use the following tools in finite algebra.
\begin{proposition}\label{FFfact1}
Let $f(x) = \prod_if_i(x)$ be a polynomial over $\mathbb{F}_2$, where $f_i(x)$ are all nonzero and pairwise relatively prime in $\mathbb{F}_2[x]$.
Then
\[\operatorname{ord}(f(x)) = \operatorname{lcm}_i\{\operatorname{ord}(f_i(x))\}.\]
\end{proposition}
\begin{proposition}\label{FFfact2}
Let $q$ be a prime power and $\alpha$ a nonzero element of the extension field $\mathbb{F}_{q^e}$ of $\mathbb{F}_q$ for a positive integer $e$.
Define $f(x) \in \mathbb{F}_q[x]$ to be the minimial polynomial of $\alpha$ over $\mathbb{F}_q$.
The order $\operatorname{ord}(f(x))$ is equal to the order of $\alpha$ in the multiplicative group $\mathbb{F}_{q^e}^*$.
\end{proposition}
For the proofs of these propositions, see \cite[Theorems 3.9 and 3.33]{Lidl:1997b}.

We now prove that a pair $\mathcal{C}_{m,t,2^h}$, $\mathcal{C}_{m,t',2^h}$ of cyclic codes achieves the trivial upper bound on
the maximum tolerable magnitude of misalignment when used as ingredients in Theorem \ref{knowntheorem}.
\begin{lemma}\label{PGorder}
Let $g(x)$ and $h(x)$ be the generator polynomials of $\mathcal{C}_{m,t,2^h}$ and $\mathcal{C}_{m,t-i,2^h}$ for a positive integer $i \leq t-1$ respectively.
Define $f(x)$ to be the quotient of $h(x) = f(x)g(x)$ divided by $g(x)$.
Then $\operatorname{ord}(f(x)) = \frac{2^{h(m+1)}-1}{2^h-1}$.
\end{lemma}
\begin{IEEEproof}
Let $\alpha$ be a primitive element in $\mathbb{F}_{2^{h(m+1)}}$ and $\beta = \alpha^{2^h-1}$.
By Theorem \ref{PGgeneratorpoly}, we have
\[f(x) = \prod_{j \in I_{m,t-i,h} \setminus I_{m,t,h}}(x-\beta^j).\]
Note that every irreducible factor of $f(x)$ over $\mathbb{F}_2[x]$ is of multiplicity $1$.
We consider two special factors of $f(x)$.
Let $j_0 = \frac{2^{h(m-t)}-1}{2^h-1}$ and $j_1 = \frac{2^{h(m-t+1)}-1}{2^h-1}-2$.
It is straightforward to see that
\begin{align*}
\max_{0 \leq i \leq h}\operatorname{w}_{2^h}(j_0(2^h-1)2^i) &= \max_{0 \leq i \leq h}\operatorname{w}_{2^h}(j_1(2^h-1)2^i)\\
&= (m-t)(2^h-1).
\end{align*}
Hence, the minimal polynomials $M_{\beta^{j_0}}(x)$, $M_{\beta^{j_1}}(x)$ of $\beta^{j_0}$ and $\beta^{j_1}$ are nonzero factors of $f(x)$.
If $\operatorname{ord}(M_{\beta^{j_0}}(x)) = \operatorname{ord}(M_{\beta^{j_1}}(x))$,
because $j_0$ and $j_1$ are relatively prime, we have
\begin{align*}
\operatorname{ord}(f(x)) &\geq \operatorname{ord}(M_{\beta^{j_0}}(x))\\
&= \frac{2^{h(m+1)}-1}{\gcd\left(2^h-1, 2^{h(m+1)}-1\right)}\\
&= \frac{2^{h(m+1)}-1}{2^h-1}.
\end{align*}
If $\operatorname{ord}(M_{\beta^{j_0}}(x)) \not= \operatorname{ord}(M_{\beta^{j_1}}(x))$,
because $M_{\beta^{j_0}}(x)$ and $M_{\beta^{j_1}}(x)$ are minimal polynomials, the two are relatively prime.
Hence, by Propositions \ref{FFfact1} and \ref{FFfact2} and the fact that $j_0$ and $j_1$ are relatively prime, we have
\begin{align*}
\operatorname{ord}(f(x)) &\geq \operatorname{lcm}\left(\operatorname{ord}\left(M_{\beta^{j_0}}(x)\right), \operatorname{ord}\left(M_{\beta^{j_1}}(x)\right)\right)\\
&= \operatorname{lcm}\left(\frac{2^{h(m+1)}-1}{\gcd\left(j_0(2^h-1), 2^{h(m+1)}-1\right)},\right.\\
&\phantom{\operatorname{lcm}\quad\quad} \left.\frac{2^{h(m+1)}-1}{\gcd\left(j_1(2^h-1), 2^{h(m+1)}-1\right)}\right)\\
&= \frac{2^{h(m+1)}-1}{2^h-1}.
\end{align*}
Since the order of a factor of the generator polynomial of a cyclic code is at most the length of the code,
in each case we have $\operatorname{ord}(f(x)) = \frac{2^{h(m+1)}-1}{2^h-1}$.
\end{IEEEproof}

We conclude this subsection with the proof of our main theorem on quantum synchronizable codes from projective geometry $\textup{PG}(m,2^h)$.
\begin{IEEEproof}[Proof of Theorem \ref{PGmain}]
Take $\mathcal{C}_{m,t,2^h}$ and $\mathcal{C}_{m,t+1,2^h}$.
By Proposition \ref{PGcyclic}, the two are both cyclic codes.
Lemma \ref{PGdualcontaining} ensures that they are dual-containing.
Lemma \ref{PGnest} guarantees that $\mathcal{C}_{m,t+1,2^h}$ is $\mathcal{C}_{m,t,2^h}$-containing.
Applying Theorem \ref{knowntheorem} with Theorem \ref{PGpara} and Lemma \ref{PGorder} proves the assertion.
\end{IEEEproof}

\subsection{Euclidean geometry codes}\label{euclidean}
This subsection investigates a different kind of finite geometry.
As in the case of projective geometry, let $m$, $h$, and $t$ be positive integers such that $t \leq m-1$.
The \textit{affine geometry} $\textup{AG}(m,2^h)$ of \textit{dimension} $m$ over $\mathbb{F}_{2^h}$ is defined as a finite geometry in which
the \textit{points} are the vectors in $\mathbb{F}_{2^h}^m$
and the $t$-\textit{dimensional subspaces} are the $t$-dimensional vector subspaces of $\mathbb{F}_{2^h}^m$ and their cosets.
Take an arbitary point in $\textup{AG}(m,2^h)$ and call it the \textit{origin}.
The \textit{Euclidean geometry} $\textup{EG}(m,2^h)$ of \textit{dimension} $m$ over $\mathbb{F}_{2^h}$
is the finite geometry obtained by deleting from $\textup{AG}(m,2^h)$ the origin and all $t$-dimensional subspaces that contain it.

Euclidean geometry has played a key role on multiple occasions in the history of coding theory.
The cyclic codes we examine here have extensively been studied in many contexts for several decades as well.
For instance, it provided a classic example of majority logic decodable codes in 1960's \cite{Chow:1967} as well as
remarkably high performance codes for modern iterative decoding \cite{Kou:2001,Tang:2005} in this century.
In quantum information theory, Euclidean cyclic codes were studied as asymmetric quantum low-density parity-check codes \cite{Sarvepalli:2009},
and quantum error-correcting codes that proved the potential of the entanglement-assisted formalism
during the last decade \cite{Hsieh:2011,Fujiwara:2010e}.
Our result on their minimum distances can be seen as a theoretical contribution to the study of this famous class of cyclic codes.

We also prove that these Euclidean cyclic codes have the best possible tolerance against misalignment
when used as the ingredients of quantum synchronizable codes.
Unfortunately, as we will show later when giving their minimum distances,
the corresponding quantum synchronizable codes can not outperform those based on primitive, narrow-sense BCH codes.
Nonetheless, because of the mathematical similarity to the case of projective geometry
as well as the importance of determining the minimum distance of a well-known cyclic code in general,
we give all mathematical details in the remainder of this section.

Let $\mathcal{B}$ be the set of $t$-dimensional subspaces of $\textup{EG}(m,2^h)$.
The incidence vector $\chi_{\pi}$ of a $t$-dimensional subspace $\pi \in \mathcal{B}$ is defined in the same way as in $\textup{PG}(m,2^h)$,
so that the entry of each coordinate is $1$ if $\pi$ contains the corresponding point and $0$ otherwise.
Define $\mathcal{E}_{m,t,2^h} = \langle \chi_\pi \mid \pi \in \mathcal{B} \rangle^\perp$
to be the dual of the vector space spanned by the incidence vectors of $t$-dimensional subspaces in $\textup{EG}(m, 2^h)$.
As in the case of projective geometry over finite fields,
the cyclic group of order $2^{hm}-1$ acts regularly on the points in the case of $\textup{EG}(m,2^h)$.
Hence, by indexing the coordinates of incidence vectors by $g^0, g^1, \dots, g^{2^{hm}-2}$ for a generator $g$ of the cyclic group in the natural order,
$\mathcal{E}_{m,t,2^h}$ can be seen as a cyclic code of length $2^{hm}-1$.
\begin{proposition}\label{EGcyclic}
$\mathcal{E}_{m,t,2^h}$ is cyclic as a linear code.
\end{proposition}

The cyclic code $\mathcal{E}_{m,t,2^h}$ is one of the oldest efficiently decodable codes.
Its basic properties in this context can be found in a modern textbook \cite[Section 13.8]{Blahut:2003}.
The following is the explicit description of the generator polynomial of $\mathcal{E}_{m,t,2^h}$.
\begin{theorem}[\cite{Chow:1967}]\label{EGgeneratorpoly}
Let $\alpha$ be a primitive element in $\mathbb{F}_{2^{hm}}$.
The generator polynomial $g(x)$ of $\mathcal{E}_{m,t,2^h}$ is
\[g(x) = \prod_{j \in I'_{m,t,h}}(x-\alpha^j),\]
where
\begin{align*}
I'_{m,t,h} = &\bigg\{ a \in \mathbb{N} \ \bigg\vert\ a \leq 2^{hm}-1,\\
&\ \left.\max_{0 \leq i \leq h}\operatorname{w}_{2^h}(a2^i) \leq (m-t)(2^h-1)\right\}.
\end{align*}
\end{theorem}

As in the case of $\mathcal{C}_{m,t,2^h}$ from $\textup{PG}(m,2^h)$,
we show that the linear code $\mathcal{E}_{m,t,2^h}$ is not only cyclic but also dual-containing and suitably nested
while having good parameters and providing maximum synchronization recovery capabilities
through Theorem \ref{knowntheorem} if $t$ is in an appropriate range with respect to $m$.

The dual-containing property is a natural consequence of a fundamental property of affine geometry.
While this fact has long been known among the finite geometry community, we give a short proof for completeness.
\begin{lemma}\label{EGdualcontaining}
If $\frac{m+1}{2} \leq t \leq m-1$, then $\mathcal{E}_{m,t,2^h}^\perp \subseteq \mathcal{E}_{m,t,2^h}$.
\end{lemma}
\begin{IEEEproof}
It suffices to prove that for any pair $\pi_0, \pi_1 \in \mathcal{B}$ of $t$-dimensional subspaces of $\textup{EG}(m, 2^h)$,
the corresponding incidence vectors $\chi_{\pi_0}, \chi_{\pi_1} \in \mathcal{E}_{m,t,2^h}^\perp$ are orthogonal to each other.
Note that $\chi_{\pi_0}$ and $\chi_{\pi_1}$ are orthogonal to each other if and only if the cardinality $\left\vert \pi_0 \cap \pi_1 \right\vert$ is even.
Because $t \geq \frac{m+1}{2}$, the intersection between $\chi_{\pi_0}$ and $\chi_{\pi_1}$ is either empty or a nonempty subspace of $\textup{AG}(m, 2^h)$.
Thus, $\left\vert \pi_0 \cap \pi_1 \right\vert$ is either $0$ or a positive integer power of $2$ as desired.
\end{IEEEproof}

The nested property of $\mathcal{E}_{m,t,2^h}$ can be shown directly through their generator polynomials.
\begin{lemma}\label{EGnested}
If $2 \leq t \leq m-1$, then $\mathcal{E}_{m,t-1,2^h} \subset \mathcal{E}_{m,t,2^h}$.
\end{lemma}
\begin{IEEEproof}
Let $a'$ be the smallest integer such that
\[\max_{0 \leq i \leq h}\operatorname{w}_{2^h}(a'2^i) = (m-t)(2^h-1)+1.\]
Then, because $2 \leq t \leq m-1$, we have
\begin{align*}
a' &= 2^{h(m-t)}+\sum_{i=0}^{m-t-1}(2^h-1)2^{hi}\\
&= 2^{h(m-t)}+2^{h(m-t-1)+1}-1\\
&< 2^{hm}-1.
\end{align*}
Hence, by Theorem \ref{EGgeneratorpoly}, the degree of the generator polynomial $g_{t-1}(x)$ of $\mathcal{E}_{m,t-1,2^h}$ is strictly larger than
that of the generator polynomial $g_t(x)$ of $\mathcal{E}_{m,t,2^h}$ while $g_t(x)$ divides $g_{t-1}(x)$.
Thus, $\mathcal{E}_{m,t-1,2^h}$ is strictly contained in $\mathcal{E}_{m,t,2^h}$.
\end{IEEEproof}

Proposition \ref{EGcyclic} and Lemmas \ref{EGdualcontaining} and \ref{EGnested}
ensure that $\mathcal{E}_{m,t,2^h}$ possesses the properties of being cyclic, dual-containing, and nested,
which are the minimum requirements in Theorem \ref{knowntheorem}.
The remainder of this section investigates the parameters of $\mathcal{E}_{m,t,2^h}$ as a code for standard error correction
and as a scheme for block synchronization recovery.

Trivially, the length of $\mathcal{E}_{m,t,2^h}$ as a linear code is exactly the number of points, which is $2^{hm}-1$.
The dimension $\dim\mathcal{E}_{m,t,2^h}$ can be directly obtained through the following formula that relates
the dimension of Euclidean geometry to that of projective geometry.
\begin{theorem}[\cite{Hamada:1968}]
For positive integers $m$, $h$, $t$ such that $t \leq m-1$,
the dimension $\dim\mathcal{E}_{m,t,2^h}^\perp$ of the vector space spanned by the incidence vectors of $t$-dimensional subspaces in $\textup{EG}(m, 2^h)$ is
\[\dim\mathcal{E}_{m,t,2^h}^\perp = \dim\mathcal{P}_{m,t,2^h}-\dim\mathcal{P}_{m-1,t,2^h}-1.\]
\end{theorem}

Since the above theorem gives the dimension of the dual,
$\dim\mathcal{E}_{m,t,2^h}$ is obtained simply by taking $2^{hm}-1-\dim\mathcal{E}_{m,t,2^h}^\perp$.
\begin{lemma}\label{EGdim}
For positive integers $m$, $h$, $t$ such that $t \leq m-1$,
the dimension of $\mathcal{E}_{m,t,2^h}$ is
\[\dim\mathcal{E}_{m,t,2^h} = 2^{hm} - \dim\mathcal{P}_{m,t,2^h}+\dim\mathcal{P}_{m-1,t,2^h}.\]
\end{lemma}
Note that the exact values of $\dim\mathcal{P}_{m,t,2^h}$ and $\dim\mathcal{P}_{m-1,t,2^h}$ can be obtained by Theorem \ref{HamadaPGdimension},
allowing for computing $\dim\mathcal{E}_{m,t,2^h}$ for given $m$, $t$, and $h$.

To prove the exact minimum distance of $\mathcal{E}_{m,t,2^h}$, we use the well-know BCH bound on the minimum distance of a cyclic code.
\begin{theorem}[BCH bound for binary codes]\label{bchbound}
Let $g(x)$ be the generator polynomial of a cyclic code of length $n$ and minimum distance $d$.
Let $n'$ be the smallest integer such that $n$ divides $2^{n'}-1$ and $\alpha$ a primitive $n$th root of unity in $\mathbb{F}_{2^{n'}}$.
If there exist a nonnegative integer $b$ and positive integer $\delta \geq 2$
such that $g(\alpha^{b+i}) = 0$ for $0 \leq i \leq \delta-2$ in $\mathbb{F}_{2^{n'}}$,
then $d \geq \delta$.
\end{theorem}

The proof of the BCH bound can be found in a standard textbook in coding theory such as \cite{MacWilliams:1977}.
We show that the BCH bound is sharp for $\mathcal{E}_{m,t,2^h}$
by explicitly constructing a codeword whose weight meets the lower bound.

A \textit{hyperoval} in a $2$-dimensional subspace of $\textup{AG}(m,2^h)$ is a set of $2^h+2$ points no three of which
are contained in the same $1$-dimensional subspace.
Such a configuration exists for all $m$ and $h$.
We show that a combination of a hyperoval and $(t-1)$-dimensional subspace
leads to a nonzero codeword of minimum weight in $\mathcal{E}_{m,t,2^h}$.
\begin{theorem}\label{EGparameters}
$\mathcal{E}_{m,t,2^h}$ is a linear $[2^{hm}-1, 2^{hm}-\dim\mathcal{P}_{m,t,2^h}+\dim\mathcal{P}_{m-1,t,2^h}, (2^{h-1}+1)2^{h(m-t-1)+1}-1]$ code.
\end{theorem}
\begin{IEEEproof}
It suffices to prove that the minimum distance $d$ of $\mathcal{E}_{m,t,2^h}$ is $(2^{h-1}+1)2^{h(m-t-1)+1}-1$.
It is straightforward to see that every positive integer $a$ smaller than $2^{h(m-t)}+2^{h(m-t-1)+1}-1$ satisfies the condition that
\[\max_{0 \leq i \leq h}\operatorname{w}_{2^h}(a2^i) \leq (m-t)(2^h-1).\]
By Theorem \ref{EGgeneratorpoly}, for all positive integer $i \leq 2^{h(m-t)}+2^{h(m-t-1)+1}-2$,
the generator polynomial $g(x)$ of $\mathcal{E}_{m,t,2^h}$ has $(x-\alpha^i)$ as its factors.
Hence, by Theorem \ref{bchbound}, we have $d \geq (2^{h-1}+1)2^{h(m-t-1)+1}-1$.
We construct a codeword of weight $(2^{h-1}+1)2^{h(m-t-1)+1}-1$.
Take a hyperoval $H$ in a $2$-dimensional subspace of $\textup{AG}(m,2^h)$.
Without loss of generality, we assume that $H$ contains the origin.
Let $L$ be the line at infinity and take a $(t-1)$-dimensional subspace $\pi$ in the hyperplane at infinity that is disjoint from $L$.
The union of $2^h+2$ parallel spaces $\{\langle p, \pi \rangle \mid p \in H\}$ is a set of $(2^h+2)2^{h(m-t-1)}$ points whose incidence vector lies in
the dual of the vector spaces spanned by the incidence vectors of $t$-dimensional subspaces of $\textup{AG}(m,2^h)$.
Removing the coordinate corresponding to the origin gives a codeword of weight $(2^h+2)2^{h(m-t-1)}-1$ in $\mathcal{E}_{m,t,2^h}$.
\end{IEEEproof}

The above theorem may be seen as a precise result on the parameters of the classic examples of majority decodable error-correcting codes
and finite geometric low-density parity-check codes as well.
As in the case of $\mathcal{C}_{m,t,2^h}$ based on the complement of projective geometry,
the exact values of all parameters for given $m$, $t$, and $h$ can be obtained by applying Theorem \ref{HamadaPGdimension} to this theorem.
Table \ref{table2} lists the parameters of $\mathcal{E}_{m,t,2^h}$ for some $m$, $t$, and $h$.
\begin{table}
\renewcommand{\arraystretch}{1.3}
\caption{Sample parameters of $\mathcal{E}_{m,t,2^h}$ from $\textup{EG}(m,2^h)$.}
\label{table2}
\centering
\begin{tabular}{rrrrrr}
\hline\hline
$m$ & $h$\rlap{\textsuperscript{a}} & $t$ & Length & Dimension & Minimum distance \\\hline
5 & 2 & 3 & 1023 & 748 & 23\phantom{0000000}\\
5 & 2 & 4 & 1023 & 988 & 5\phantom{0000000}\\
5 & 3 & 3 & 32767 & 28042 & 79\phantom{0000000}\\
5 & 3 & 4 & 32767 & 32552 & 9\phantom{0000000}\\
6 & 2 & 4 & 4095 & 3572 & 23\phantom{0000000}\\
6 & 2 & 5 & 4095 & 4047 & 5\phantom{0000000}\\
6 & 3 & 4 & 262143 & 249816 & 79\phantom{0000000}\\
6 & 3 & 5 & 262143 & 261801 & 9\phantom{0000000}\\
6 & 4 & 4 & 16777215 & 16490000 & 287\phantom{0000000}\\
6 & 4 & 5 & 16777215 & 16774815 & 17\phantom{0000000}\\
7 & 2 & 4 & 16383 & 11728 & 95\phantom{0000000}\\
7 & 2 & 5 & 16383 & 15473 & 23\phantom{0000000}\\
7 & 2 & 6 & 16383 & 16320 & 5\phantom{0000000}\\
7 & 3 & 4 & 2097151 & 1763104 & 639\phantom{0000000}\\
7 & 3 & 5 & 2097151 & 2068983 & 79\phantom{0000000}\\
7 & 3 & 6 & 2097151 & 2096640 & 9\phantom{0000000}\\
8 & 2 & 5 & 65535 & 55627 & 95\phantom{0000000}\\
8 & 2 & 6 & 65535 & 64055 & 23\phantom{0000000}\\
8 & 2 & 7 & 65535 & 65455 & 5\phantom{0000000}\\
8 & 3 & 5 & 16777215 & 15742657 & 639\phantom{0000000}\\
8 & 3 & 6 & 16777215 & 16719003 & 79\phantom{0000000}\\
8 & 3 & 7 & 16777215 & 16776487 & 9\phantom{0000000}\\
9 & 2 & 5 & 262143 & 184848 & 383\phantom{0000000}\\
9 & 2 & 6 & 262143 & 242724 & 95\phantom{0000000}\\
9 & 2 & 7 & 262143 & 259860 & 23\phantom{0000000}\\
9 & 2 & 8 & 262143 & 262044 & 5\phantom{0000000}\\
\hline\hline
\multicolumn{6}{l}{\scriptsize\textsuperscript{a}
When $h=1$, the parameters of $\mathcal{E}_{m,t,2}$ and $\mathcal{C}_{m-1,t,2}$ coincide.}
\end{tabular}
\end{table}

It should be noted that, unlike the projective case, the length of this type of cyclic code is always a Mersenne number.
When compared to the primitive, narrow-sense BCH code of the same length,
the generator polynomial of $\mathcal{E}_{m,t,2^h}$ is of higher degree, which implies that $\mathcal{E}_{m,t,2^h}$ has a smaller dimension.
Now the BCH bound is a lower bound on the minimum distance of the BCH code.
However, the above theorem shows that the minimum distance of $\mathcal{E}_{m,t,2^h}$ actually matches the BCH bound.
Hence, the cyclic codes based on Euclidean geometry are generally poorer in terms of information rate and minimum distance.
Considering the decent error-correcting performance reported in the literature,
the poorer minimum distance property seems to suggests that the error-correcting codes based on Euclidean geometry
benefit more from decoding algorithms that are less sensitive to true minimum distances
such as the sum-product algorithm and majority logic algorithm.

Now, in the context of quantum synchronizable coding, we would like as high misalignment tolerance as possible.
As is the case with projective geometry, we prove that any pair of cyclic codes taken from
the nested chain $\mathcal{E}_{m,\lceil\frac{m+1}{2}\rceil,2^h} \subset  \mathcal{E}_{m,\lceil\frac{m+1}{2}\rceil+1,2^h} \subset \dots \subset
\mathcal{E}_{m,m-1,2^h}$ attains the trivial upper bound in this regard.
\begin{lemma}\label{EGorder}
Let $g(x)$ and $h(x)$ be the generator polynomials of $\mathcal{E}_{m,t,2^h}$ and $\mathcal{E}_{m,t-i,2^h}$ for a positive integer $i \leq t-1$ respectively.
Define $f(x)$ to be the quotient of $h(x) = f(x)g(x)$ divided by $g(x)$.
Then $\operatorname{ord}(f(x)) =2^{hm}-1$.
\end{lemma}
\begin{IEEEproof}
By Theorem \ref{EGgeneratorpoly}, we have
\[f(x) = \prod_{j \in I'_{m,t-i,h} \setminus I'_{m,t,h}}(x-\alpha^j).\]
Let $j_0 = 2^{h(m-t)}-1$ and $j_1 = 2^{h(m-t)}-2$. It is easy to see that these two relatively prime integers are in the set $I'_{m,t-i,h} \setminus I'_{m,t,h}$.
Write the minimal polynomials of $\alpha^{j_0}$ and $\alpha^{j_1}$ as $M_{\alpha^{j_0}}(x)$ and $M_{\alpha^{j_1}}(x)$ respectively.
By Propositions \ref{FFfact1} and \ref{FFfact2}, we have
\begin{align*}
\operatorname{ord}(f(x)) &\geq \operatorname{lcm}\left(\operatorname{ord}\left(M_{\alpha^{j_0}}(x)\right), \operatorname{ord}\left(M_{\alpha^{j_1}}(x)\right)\right)\\
&= \operatorname{lcm}\left(\frac{2^{hm}-1}{\gcd\left(j_0, 2^{hm}-1\right)}, \frac{2^{hm}-1}{\gcd\left(j_1, 2^{hm}-1\right)}\right)\\
&= 2^{hm}-1.
\end{align*}
Since the order of a factor of the generator polynomial of a cyclic code is at most the length of the code,
we have $\operatorname{ord}(f(x)) = 2^{hm}-1$ as desired.
\end{IEEEproof}

The following theorem summarizes the results presented in this subsection.
\begin{theorem}\label{EGmain}
Let $m$, $h$, and $t$ be positive integers such that $\frac{m+1}{2} \leq t \leq m-2$.
For every pair $a_l$, $a_r$ of nonnegative integers such that $a_l + a_r < 2^{hm}-1$
there exists a quantum synchronizable
$(a_l, a_r)$-$\bigl[\bigl[2^{hm}-1+a_l+a_r, 2^{hm}-2\dim\mathcal{P}_{m,t,2^h}+2\dim\mathcal{P}_{m-1,t,2^h}+1\bigr]\bigr]$ code that corrects
at least up to $(2^{h-1}+1)2^{h(m-t-1)}-1$ phase errors and
at least up to $(2^{h-1}+1)2^{h(m-t-2)}-1$ bit errors.
\end{theorem}
\begin{IEEEproof}
Apply Theorem \ref{knowntheorem} to $\mathcal{E}_{m,t,2^h}$ and $\mathcal{E}_{m,t+1,2^h}$
with Proposition \ref{EGcyclic}, Lemmas \ref{EGdualcontaining}, \ref{EGnested}, and \ref{EGorder}, and Theorem \ref{EGparameters}.
A routine calculation proves the assertion.
\end{IEEEproof}

\section{Concluding remarks}\label{conclusion}
We constructed a family of quantum synchronizable codes that correct both standard quantum errors and block synchronization errors.
One type of our code exploits the complement structure of projective geometry while the other type takes direct advantage of Euclidean geometry without the origin.
Both types of codes are proved to achieve the highest possible tolerance against misalignment.
The results presented in this paper enrich the variety of available quantum synchronizable error-correcting codes.

While cyclic codes are useful in both classical and quantum information theory,
it is not easy to construct ones with large minimum distances.
A particularly difficult task is to precisely determine the parameters instead of bounding them from above or below.
For instance, given a generator polynomial, it is a very challenging algebraic problem to give the exact minimum distance of the corresponding cyclic code.
In fact, precise results on cyclic codes with fairly large minimum distances such as Theorems \ref{PGpara} and \ref{EGparameters}
given in this paper are not common in the literature.
Significant mathematical advances in this aspect are much desired.

It is also notable that the proof of Theorem \ref{PGpara} gives a complete picture of the minimum weight nonzero codewords
of our cyclic codes $\mathcal{C}_{m,t,2^h}$ based on projective geometry.
In fact, its minimum weight nonzero codewords all come from the incidence vectors of $(m-t)$-spaces in $\textup{PG}(m,2^h)$.

In the case of $\mathcal{E}_{m,t,2^h}$, however, it appears more difficult to obtain a similar classification of the minimum weight nonzero codewords.
Although we did not identify all codewords of weight $(2^{hm}+2)2^{m-t-1}-1$,
we conjecture that every nonzero codeword of minimum weight in $\mathcal{E}_{m,t,2^h}$
is obtained in the way shown in the proof of Theorem \ref{EGparameters}.

Another notable point regarding our constructions for quantum synchronizable codes is that
while Theorems \ref{PGmain} and \ref{EGmain} were proved by using a pair of cyclic codes lying next to each other in a chain of nested codes,
we can employ any pair, such as $\mathcal{C}_{m,t,2^h}$ and $\mathcal{C}_{m,t+2,2^h}$, in the same chain to obtain analogous theorems.
The resulting quantum synchronizable codes will have the same highest possible tolerance against misalignment,
the same length, the same dimension, and the same phase error correction capabilities as in Theorems \ref{PGmain} and \ref{EGmain}.
The advantage is that these alternative codes require fewer quantum interactions for detecting bit errors
because fewer stabilizer operators are involved.
They have a drawback of reduced bit error correction capabilities because
the cyclic codes responsible for bit error detection will have smaller minimum distances.

Note that the frameworks of quantum synchronizable codes given in \cite{Fujiwara:2013d,Fujiwara:2013}
implicitly assume that phase errors are more likely,
which is a reasonable assumption because asymmetry in error probability between bit errors and phase errors
is expected in actual quantum devices \cite{Ioffe:2007}.
Our main results also put emphasis more on the minimum distance responsible for phase error correction than that for bit error correction.
However, in a situation where the quantum channel is very highly asymmetric
and introduces phase errors far more frequently,
using even more asymmetric quantum synchronizable error-correcting codes may make more sense than employing
the fairly asymmetric ones given in Theorems \ref{PGmain} and \ref{EGmain}.
We hope that our results presented in this paper help advance the field in various directions from both mathematical and physical viewpoints.

\section*{Acknowledgment}
The authors thank the anonymous reviewers
and Associate Editor Alexei Ashikhmin for careful reading of the manuscript and constructive suggestions.



\begin{IEEEbiographynophoto}{Yuichiro Fujiwara}
(M'10) received the B.S. and M.S. degrees in mathematics from Keio University, Japan,
and the Ph.D. degree in information science from Nagoya University, Japan.

He was a JSPS postdoctoral research fellow with the Graduate School of
System and Information Engineering, Tsukuba University, Japan, and a visiting
scholar with the Department of Mathematical Sciences, Michigan Technological University.
He is currently with the Division of Physics, Mathematics and Astronomy,
California Institute of Technology, Pasadena, where he works as a JSPS postdoctoral research fellow.

Dr.\ Fujiwara's research interests include combinatorics and its interaction with computer science, quantum information science, and electrical engineering,
with particular emphasis on combinatorial design theory, algebraic coding theory, and quantum information theory.
\end{IEEEbiographynophoto}

\begin{IEEEbiographynophoto}{Peter Vandendriessche}
(S'11--M'14) simultaneously received the M.Sc. in Mathematics and the M.Sc. in Mathematical Informatics in 2010 at Ghent university, Belgium.
Since then he has been working at Ghent University, where he received the Ph.D. degree in Mathematics in 2014.
He is currently supported by a PhD fellowship of the Research Foundation - Flanders (FWO).
\end{IEEEbiographynophoto}

\end{document}